# Isolated electron spins in silicon carbide with millisecond-coherence times

David J. Christle[1,2], Abram L. Falk[1], Paolo Andrich[1,2], Paul V. Klimov[1,2], Jawad ul Hassan[3], Nguyen T. Son[3], Erik Janzén[3], Takeshi Ohshima[4], and David D. Awschalom[1,2,‡]

[1] Inst. for Molecular Engineering, University of Chicago, Chicago, IL 60637, USA
[2] Dept. of Physics, University of California, Santa Barbara, Santa Barbara, CA 93106, USA
[3] Dept. of Physics, Chemistry and Biology, Linköping University, SE-58183 Linköping, Sweden
[4] Japan Atomic Energy Agency, 1233 Watanuki, Takasaki, Gunma 370-1292, Japan

[‡] Email address: awsch@uchicago.edu

**The elimination of defects from SiC has facilitated its move to the forefront of the optoelectronics and power-electronics industries[1]. Nonetheless, because the electronic states of SiC defects can have sharp optical and spin transitions, they are increasingly recognized as a valuable resource for quantum-information and nanoscale-sensing applications[2-16]. Here, we show that individual electron spin states in highly purified monocrystalline 4H-SiC can be isolated and coherently controlled. Bound to neutral divacancy defects[2,3], these states exhibit exceptionally long ensemble Hahn-echo spin coherence, exceeding 1 ms. Coherent control of single spins in a material amenable to advanced growth and microfabrication techniques is an exciting route to wafer-scale quantum technologies.**

Isolated electron spins are a promising platform for an array of new technologies, ranging from quantum communication[17] to nanoscale nuclear magnetic resonance[18,19] and intracellular sensing of magnetic, electric and thermal fields[20,21]. By exploiting spin-dependent optical transitions, optically detected magnetic resonance (ODMR) has proven to be a powerful technique for achieving single-spin addressability in the solid state[22-24]. As with the diamond nitrogen-vacancy (NV) center[17-21,23], a particular focus of such research, the neutral divacancy in SiC has a spin-triplet electronic ground state[2,3] that can be polarized and read out with ODMR[6,8,11,13,25-27]. However, ODMR has previously only been used in SiC to measure spin ensembles, not individual electronic states. Here, by showing that neutral divacancies in SiC are individually addressable and have outstanding Hahn-echo coherence times ($T_2$ times), our results open up many new avenues for engineering high-performance electronic devices incorporating single-spin sensors and memories.

Achieving single-center addressability with ODMR requires a nearly defect-free substrate, such that multiple defects do not occupy a diffraction-limited confocal volume. Towards this end, we use hot-wall chemical vapor deposition[28] to grow a 120-micron-thick single-crystal epitaxial film on an n-type 4H-SiC substrate. This growth technique can be used to create commercial quality, multilayer electronic structures at the wafer scale. Our epilayer is optimized to have no dislocations or polytype inclusions, and in addition, a very low ($5 \times 10^{13}$ $cm^{-3}$) unintentional dopant density. After mechanically separating the epilayer from the substrate, we polish and dice it, and then irradiate the diced samples with 2 MeV electrons at a range of fluences ($5 \times 10^{12}$ $cm^{-2}$ to $1 \times 10^{15}$ $cm^{-2}$) in order to create Si and C vacancies. Finally, we anneal the samples to activate vacancy migration, whereby divacancies are formed[11,27].

To measure the photoluminescence (PL) from single divacancies, we integrate a high efficiency superconducting nanowire single photon detector into a home-built confocal microscopy setup that uses a 975 nm continuous-wave excitation laser. With the sample at 20 K, we observe distinct bright spots (3-5 kcounts/s) in a scanning PL image (Fig. 1a). We then use Hanbury-Brown Twiss interferometry to measure the second-order intensity correlation function ($g^{(2)}$) of the emission from several of these spots, all of which display photon anti-bunching behavior to varying degrees. The three divacancies whose spin measured for this work exhibit $g^{(2)}(t = 0) < 0.5$, where $t$ is the time delay between successive photons, indicating that they are single quantum emitters (Fig. 1b). The characteristic times of the $g^{(2)}(t)$ dips range from 9 to 12 ns, slightly less than the neutral divacancies' (14 ± 3 ns) optical lifetimes[13], as expected from an optically pumped emitter.

Because there are two inequivalent lattice sites for C and Si atoms in 4H-SiC, namely the hexagonal (*h*) and quasi-cubic (*k*) sites, four inequivalent neutral divacancies can exist in 4H-SiC. We collect complete anti-bunching and ODMR data from three of these: the (*hh*), (*kk*), and (*kh*) divacancy forms (Fig. 1c). We do not observe the (*hk*) form using either ensemble (see SI) or confocal PL measurements, suggesting that it is less stable or has a higher formation energy.

We measure single-spin ODMR spectra (Fig. 2a) by sweeping the frequency ($f$) of an applied microwave driving field and measuring the change in PL from individual divacancies. Spin-selective transitions cause the emitted PL intensity to depend on the divacancy's spin-sublevel occupation[6,8,11]. Therefore, the PL intensity has a peak or a dip when $f$ is resonant with a spin transition.

Each inequivalent divacancy has a characteristic spin-resonance signature. For the c-axis-oriented (*hh*) and (*kk*) divacancies, we measure ODMR with an applied c-axis oriented magnetic field ($B$) to lift the degeneracy of the $m_s = \pm 1$ spin transitions. For the (*kk*) divacancy that we measure, each resonance line was split by (13.4 ± 0.2) MHz, likely due to hyperfine coupling to a nearby $^{29}Si$ nucleus[3]. For the (*kh*) defect, we measure ODMR at $B = 0$, because the crystal field already splits the degeneracy between all three spin sublevels of the basal-plane-oriented defects.

In addition to continuous-wave ODMR measurements, we observe coherent Rabi oscillations of the single divacancies by applying variable-length bursts of resonant microwaves between short initialization and readout laser pulses (Fig. 2b). These oscillations are the simplest demonstration of coherent control of the spin within a two-level subspace of its spin-1 ground state.

A long $T_2$ time is important for both sensing and quantum-information applications of a solid-state spin. To characterize $T_2$, we measure the neutral divacancies' Hahn-echo spin coherence ($C$) as a function of free precession time ($t_{free}$), using ensembles of defects in order to expedite data acquisition (Fig. 3). The spin coherence is characterized by collapses and revivals with $t_{free}$, an effect known as electron spin echo envelope modulation (ESEEM)[29]. The ESEEM oscillations originate from periodic dephasing and rephasing due to the Larmor precession of naturally abundant, spin-1/2 $^{29}$Si and $^{13}$C nuclei in the sample, with the dominant 58 kHz ESEEM oscillation (Fig. 3, inset) due to the more abundant $^{29}$Si nuclei. The (1.25 ± 0.05) ms fitted $T_2$ time is significantly longer than the previously reported 360 μs $T_2$ time measured for divacancies in SiC substrates with higher defect densities[11]. Moreover, although $^{29}$Si is more abundant than $^{13}$C, (4.7% vs. 1.1% natural abundance), this $T_2$ time is twice as long as the 600 μs $T_2$ times measured for NV centers in chemically but not isotopically purified diamond[30]. Isotopic purification and dynamical decoupling sequences will likely increase this $T_2$ time even further.

Incorporating these highly coherent single electron spins into high-performance SiC devices should provide many new opportunities for advancing quantum control. In the future, spin-photon entanglement in SiC could offer a promising route towards quantum-repeater networks, facilitated by the divacancy emission near telecom wavelengths. Moreover, spins embedded into SiC transistors could lead to electrically gated spin-spin coupling via charge-state manipulation, while spins within high-Q SiC micromechanical resonators could be a platform for studying spin-phonon interactions. Just as the performance of commercial SiC electronics has been improved by a greater understanding of defect science, future avenues for defect-based quantum technologies may be driven by SiC devices.

# Methods

## Sample preparation

A 120-μm-thick epilayer of single-crystal 4H-SiC was grown on an on-axis 4H-SiC substrate[28]. The samples were irradiated at room temperature with 2 MeV electrons from a Van de Graaf generator, creating Si and C vacancies. A 30-minute long, 750 °C anneal in Ar gas caused the vacancies to migrate and form vacancy complexes, including divacancies[27]. The diced silicon carbide samples were cleaned with acetone and isopropanol, followed by a deionized water rinse. Remaining surface contaminants were removed using a 3:1 mixture of concentrated sulfuric acid and hydrogen peroxide heated to 90 °C for 30 minutes.

## Near-infrared confocal ODMR

All experiments were performed in a Janis ST-500 liquid helium flow cryostat at a temperature of 20 K. A 975-nm laser was used to non-resonantly excite the divacancy defects through their phonon absorption sidebands. The SiC membranes were mounted on top of an antenna consisting of a short-terminated coplanar waveguide made from patterned Au on a duroid substrate. Our home-built confocal microscope used a 0.85 NA 100x near-infrared objective (Olympus, LCPLN100XIR) to focus the excitation light and to collect the emitted PL. The excitation power was 1 mW at the back of the objective. The PL was filtered to collect only between 1.1 μm and 1.6 μm and was then focused into a single mode fiber that served as the confocal pinhole. A commercial closed-cycle NbTiN superconducting nanowire single photon detector (SingleQuantum, EOS) with approximately 28% quantum efficiency was used to register the near-infrared photons, and the antibunching measurements used a time-correlated photon counting card (PicoQuant, PicoHarp PH300) to collect the conditional photon statistics.

The GHz microwaves used for spin resonance experiments were generated by a signal generator (National Instruments, PXIe-5652) and then amplified (MiniCircuits, ZHL-30W-252) before reaching the antenna.

### Data analysis

The measurements of continuous-wave spin resonance and Rabi oscillations are background-corrected by subtracting the PL from the substrate outside of the confocal spot. This background consists of roughly 40% of the PL signal. The antibunching measurements are rebinned to a larger bin size for visual clarity, but the inferred depth of the dip from a fit to these data can be slightly skewed as an artifact of this process. To mitigate this artifact, we derived the plotted fits from a Bayesian approach applied to the raw data alone (see SI). This technique takes into account the Poisson error in each bin and infers only estimates and error bars that are physically realizable ($g^2(0) \geq 0$). These measurements are also background-corrected, where the background is measured about 1.2 microns away from the center of the PL.

## Acknowledgments

The authors thank Ádám Gali, Bob Buckley, William Koehl, F. Joseph Heremans, and Greg Calusine for helpful discussions. The authors also thank Sergey Chemerisov and Norstel AB for assistance preparing preliminary samples and gratefully acknowledge support from the NSF, AFOSR MURI, the Knut & Alice Wallenberg Foundation, the Linköping Linnaeus Initiative for Novel Functionalized Materials, and the Swedish Government Strategic Research Area Grant in Materials Science (Advanced Functional Materials).

## Author Information

### Contributions

J.H., E.J., and N.T.S. contributed to design, growth and processing of the SiC samples. T.O. and N.T.S contributed to electron irradiation and annealing experiments. D.J.C., A.L.F., P.A., and P.V.K performed the optical experiments. All of the authors contributed to the analysis of the data, discussions and the production of the manuscript.

**Competing Financial Interests:** The authors declare no competing financial interests

**Corresponding author:** Correspondence should be addressed to David D. Awschalom: awsch@uchicago.edu.

## Figures and captions:

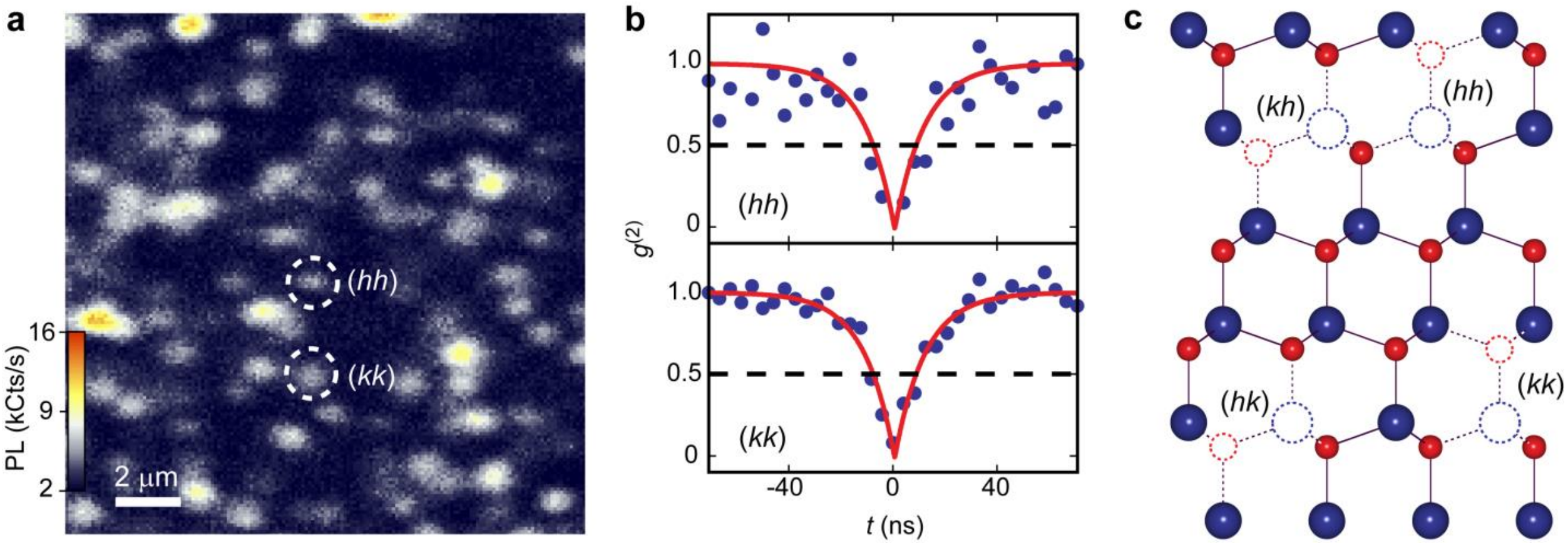

**Figure 1. Isolation of neutral divacancies in SiC. a.** A 16μm x 16 μm confocal PL image from a 4H-SiC membrane irradiated at $10^{13}$ $cm^{-2}$ fluence. Confocal PL is collected at a depth of 25 μm into the membrane, and the sample temperature is held at 20 K. **b.** $g^{(2)}(t)$ measurements for the two circled divacancies in Fig. 1a, whose forms are the (*hh*) divacancy (top) and the (*kk*) divacancy (bottom). The $g^{(2)}$ curves (blue dots) show strong antibunching, clearly achieving the $g^{(2)}$ $(t = 0) < 0.5$ threshold for single optical emitters. The red curves are fits to a simple two-level model. Fit details and antibunching data for a third divacancy (the (*kh*) form, outside of the scan in Fig. 1a) are presented in the SI. **c.** Divacancies in 4H-SiC consist of neighboring Si and C vacancies. Because either the *h* or *k* lattice site can be vacant, there are four inequivalent forms of divacancy in 4H-SiC.

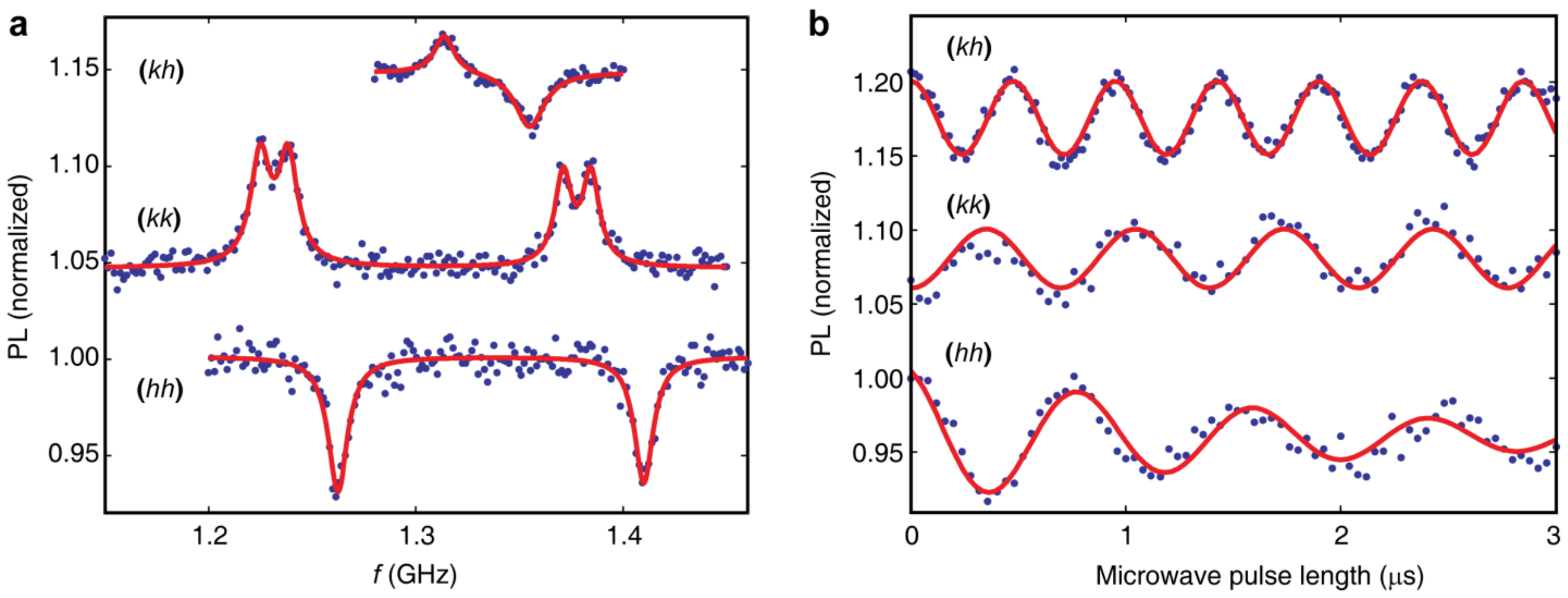

**Figure 2. Coherent control of single divacancy spins. a.** Measurements of continuous-wave ODMR (blue dots). The measured (*hh*) and (*kk*) divacancies are circled in Fig. 1a. For these two divacancies, a c-axis oriented $B$ = 50 G is applied when measuring these defects to lift the $m_s$ = ±1 degeneracy at $B$ = 0. For the (*kh*) divacancy, no external $B$ is applied. The temperature is 20 K. The three curves are vertically offset, for clarity, and the red curves are fits to Lorentzians. **b**. Rabi oscillations (blue dots) of the three divacancies measured in Fig. 2a, demonstrating coherent control of single electron spins in SiC. The fits (red curves) are to decaying sinusoids.

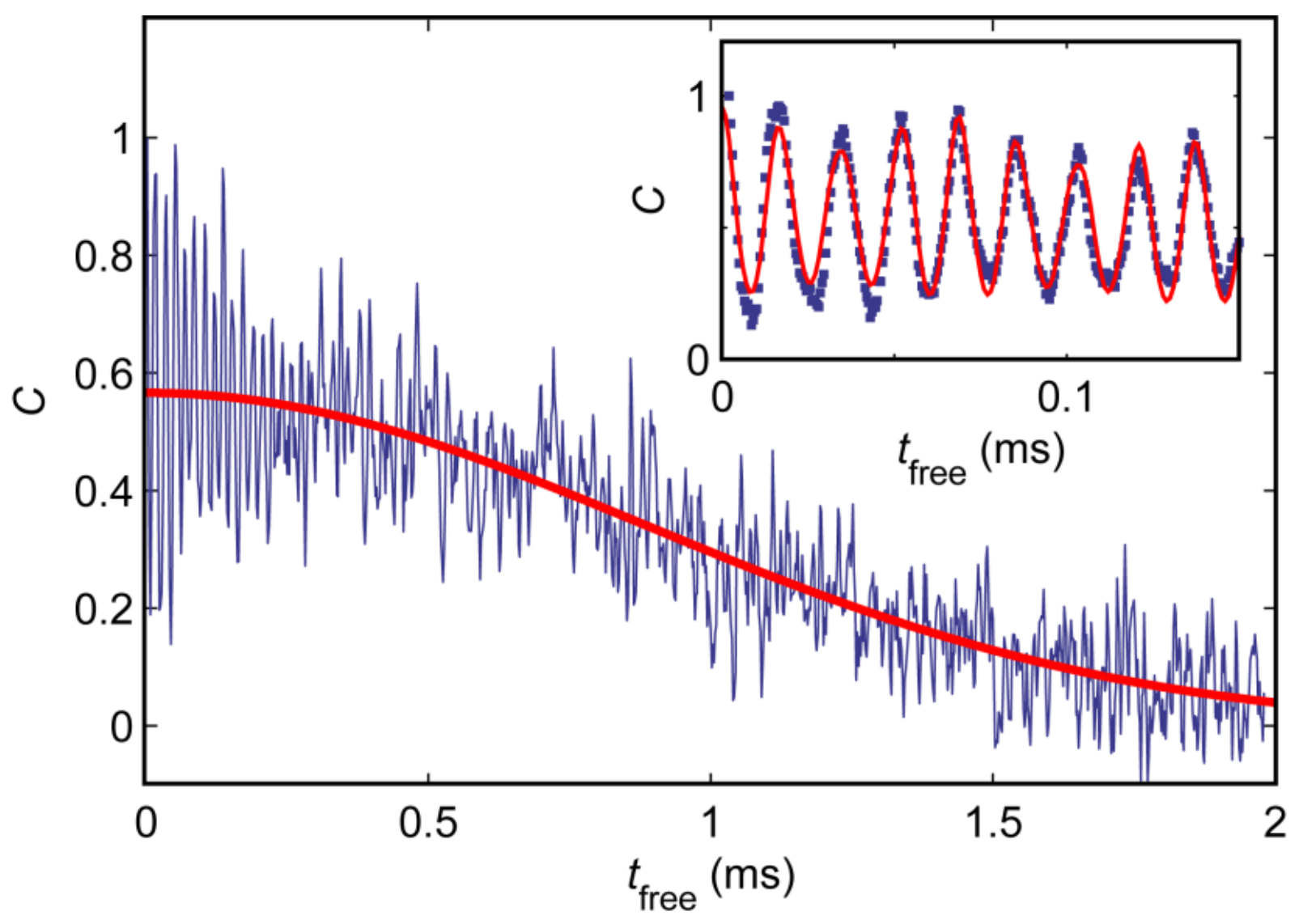

**Figure 3. Hahn-echo spin coherence.** Measurement of ensemble $C(t_{\mathrm{free}})$ of the ($kk$) divacancy from the sample irradiated at $10^{15}$ cm$^{-2}$ fluence. For these data, $B$ = 135 G and the sample temperature is 20 K (blue curve). The fit (red curve) of $C(t_{\mathrm{free}})$ is to a compressed exponential of the form $e^{-(t_{\mathrm{free}}/T_2)^n}$, where $T_2$ and $n$ are free parameters. The fit yields $T_2 = (1.25 \pm 0.05)$ ms and $n = (2.0 \pm 0.2)$. The ESEEM oscillations are due to periodic spin dephasing from the spin bath of naturally abundant, spin-1/2 $^{29}$Si and $^{13}$C nuclei, which are precessing at their Larmor frequencies (58 kHz and 73 kHz respectively). The ESEEM oscillations exceed the noise when $t_{\mathrm{free}}$ is less than roughly 0.5 ms. Inset: Hahn-echo spin coherence measured over a shorter interval, showing the ESEEM oscillations more clearly. The dominant oscillation is due to $^{29}$Si nuclear precession. The fit is discussed in the SI and in Ref. 29.